\documentclass[11pt,a4paper]{article}
\usepackage{graphicx,url,amsmath,amssymb,alltt}
\usepackage{multibib}
\newcites{m}{Bibliography}
\title{Michael John Caldwell Gordon (FRS 1994)\\  28 February 1948 -- 22 August 2017}
\author{Lawrence C Paulson FRS\\
       Computer Laboratory, University of Cambridge\\ \texttt{lp15@cam.ac.uk}}
\let\ts=\thinspace
\begin{document}
\maketitle
\begin{abstract}
	 Michael Gordon was a pioneer in the field of interactive theorem proving and hardware verification. In the 1970s, he had the vision of formally verifying system designs, proving their correctness using mathematics and logic. He demonstrated his ideas on real-world computer designs. His students extended the work to such diverse areas as the verification of floating-point algorithms, the verification of probabilistic algorithms and the verified translation of source code to correct machine code. He was elected to the Royal Society in 1994, and he continued to produce outstanding research until retirement.

His achievements include his work at Edinburgh University helping to create Edinburgh LCF, the first interactive theorem prover of its kind, and the ML family of functional programming languages. He adopted higher-order logic as a general formalism for verification, showing that it could specify hardware designs from the gate level right up to the processor level. It turned out to be an ideal formalism for many problems in computer science and mathematics. His tools and techniques have exerted a huge influence across the field of formal verification.
\end{abstract}

%

\section{Early Life}

Mike Gordon was born in Ripon, Yorkshire to John Gordon and Daphne Mavis Gordon (n\'ee More). He had perhaps a lonely childhood: he was an only child, and his father committed suicide when Mike was eight years old. His mother sent him as a boarding pupil first to ``the notorious Dartington Hall'' (where he forgot how to read) and then to Bedales school, which he regarded ``as being my home between the ages of eight and 18'' \citem{gordon-fifty}. Bedales was then a mixed, progressive school specialising in the arts.

Mike was a quiet pupil but showed early signs of a lively, scientific mind. He built model aeroplanes, some petrol powered and radio controlled. Once he slipped into the chemistry lab and synthesised methyl mercaptan, to impress his friends with its terrible smell. On another occasion, he made nitrogen triiodide crystals --- which explode when stepped on --- and sprinkled them in the library.\footnote{Simon Laughlin, email, 8 February 2018} Pupils called him Gecko because of his bright, prominent eyes and surprised expression: a look he never lost.\footnote{Stephen Levinson, email, 17 January 2018}

In 1966, Mike was accepted to Cambridge University to study engineering. As preparation, he took a gap year as a management trainee at the North Thames Gas Board~\citem{gordon-management}. This was his first exposure to the real world after a childhood spent at boarding school, and it came as a shock. The staff were divided on class lines, white coats for the management and brown coats for the workers, with separate toilets and canteens. He observed time and motion studies and the compilation of tables listing, for example, ``how long it would take to put a single screw into a wall for different screw sizes''; these data would then be used to set deadlines for workers. He liked to joke about this system, but he clearly saw it as wasteful and oppressive. He spent much of his time at the Beckton Gas Works: a vast, bleak and partly derelict site that would later become the shattered city of Hu{\'e} in Stanley Kubrick's Vietnam war movie, \textit{Full Metal Jacket}.

Mike's gap year experience destroyed his enthusiasm for engineering. But during this time he stumbled upon symbolic logic, buying logic books to read while commuting between home and the Beckton Gas Works. And so he decided to study mathematics as ``the furthest subject from engineering that didn't involve writing essays''.  Initially he struggled with mathematics (his subject change would be forbidden today), but he improved year after year and eventually graduated with a First:
\begin{quote}
Although I found the course very tough, it gave me the tools and confidence to feel that with sufficient effort \ldots\ I could master any mathematical material I needed. This laid a solid foundation for my subsequent academic career. \citem{gordon-struggling}
\end{quote}

Mike's first exposure to computers came in 1969, after his second year at Cambridge, when he took a summer job at the National Physical Laboratory (NPL)~\citem{gordon-npl}. He learnt how to boot up a Honeywell DDP-516 by manually keying in a loader using switches and to load machine code via paper tape.  This machine was likely the inspiration for the 16-bit minicomputer that Mike designed later as the canonical example for his verification techniques. He worked on pattern recognition, writing code to identify printed characters by testing for specific features. Today, machine learning is invariably used for such tasks, and in fact Mike wrote a final year essay on perceptrons, a primitive type of neural network. This experience lured Mike to Edinburgh University's School of Artificial Intelligence, where he ultimately specialised in programming language theory.
 
\section{Research Milieu: Verification and Semantics}

Computer programming has been plagued by errors from the earliest days. Ideas for verifying programs mathematically proliferated during the 1960s. Robert Floyd proposed a methodology for attaching and verifying logical assertions within flowcharts~\cite{floyd-assigning}. In a landmark paper~\cite{hoare-axiomatic}, C A R Hoare (FRS 1982) proposed a similar technique but taking the form of a novel logical calculus combining program fragments and mathematical assertions. It worked beautifully, at least on small examples.

This technique was a form of \textit{programming language semantics}: a precise specification of the meaning of every construct of a given programming language. For example, consider the program fragment \texttt{A+B}, for computing the sum of the values of \texttt{A} and~\texttt{B}, two computable expressions. What happens if the sum is too large to be represented on the computer? What if \texttt{B}, although nonzero, is much smaller than~\texttt{A}, so precision is lost and \texttt{A+B} turns out to equal~\texttt{A}? Further complications arise if evaluating \texttt{A} and \texttt{B} causes side effects, such as writing to memory; then there is no reason why \texttt{A+B} should equal \texttt{B+A} or why \texttt{A+A} should equal \texttt{2*A}. For another example, suppose we have a vector \texttt{V} whose components are \texttt{V[1]}, \ldots, \texttt{V[$n$]}, and consider a command to copy data into~\texttt{V}\@. If more than $n$ elements are supplied then they may get copied into an arbitrary part of memory. This is the classic buffer overflow error, which has caused innumerable security vulnerabilities. One remedy for such issues is to precisely specify the semantics of every programming language construct so that ambiguities and vulnerabilities can be identified and eliminated.

During the 1960s, Dana Scott and Christopher Strachey were developing the \textit{denotational} approach to semantics~\cite{scott-outline}. This involves defining functions mapping programming constructs such as expressions, statements and types into suitable mathematical domains. A key idea is the use of \textit{partial orderings} to deal with non-termination. For example, if $f$ and $g$ are computable partial functions on the natural numbers, then $f\sqsubseteq g$ means that for all~$x$, if $f(x)$ is defined then $g(x)=f(x)$, and we say ``$f$ approximates~$g$''. That idea came from recursive function theory. But once we accept that not everything is a number and grasp the need for functions themselves to be values, this simplifies to $f\sqsubseteq g$ if and only if $f(x)\sqsubseteq g(x)$ for all~$x$. Basic domains like the natural numbers are made into partial orderings by affixing a ``bottom element''~$\bot$, with $\bot\sqsubseteq n$ for every natural number~$n$. Domain theory requires functions to be \textit{monotonic} --- if $x\sqsubseteq y$ then $f(x)\sqsubseteq f(y)$. The intuition is that a computable function cannot know that its argument is failing to terminate, and can never do more with less. Functions must also be \textit{continuous} (limit-preserving). The intuition is that an infinite computation delivers nothing more than the results of successive finite computations. Sometimes called \textit{fixed-point theory}, these techniques could specify the semantics of any recursive function definition. 

Scott's 1970 Oxford technical report~\cite{scott-outline} ---  still rewarding to read --- outlined this mathematically sophisticated and elegant approach. It set off a frenzy of activity. Researchers strove to extend and simplify Scott and Strachey's highly abstruse techniques, while relating them to Hoare logic on the one hand and to more intuitive semantic notions on the other. 


Denotational semantics makes heavy use of the \textit{$\lambda$-calculus} \cite{barendregt}: a tiny, primitive language of functions. Terms of the $\lambda$-calculus include 
\begin{itemize}
	\item \emph{variables}~$x$, $y$, $z$,~$\ldots$
	\item \textit{abstractions} $(\lambda x.M)$, where $M$ is a term, and
	\item \textit{applications} $(MN)$, where $M$ and $N$ are terms.
\end{itemize}
The abstraction $(\lambda x.M)$ is intended to represent a
function, and $((\lambda x.M)N)$ can be ``reduced'' to $M[N/x]$: the result of substituting $N$ for~$x$ in~$M$. Versions of the $\lambda$-calculus are used in denotational semantics and higher-order logic. The original, \textit{untyped} $\lambda$-calculus can express arbitrary computations, but its terms are meaningless symbol strings. The \textit{typed} $\lambda$-calculus assigns types to all variables, yielding a straightforward set-theoretic semantics: types denote sets and abstractions denote functions. The typed system is therefore more intuitive, but also more restrictive. It assigns $(\lambda x.M)$ the type $\sigma\to\tau$ if $x$ has type $\sigma$ and $M$ has type $\tau$; it allows $(MN)$ only if $M$ has type $\sigma\to\tau$ and $N$ has type $\sigma$. It rejects terms like $(\lambda xy.y(xxy))(\lambda xy.y(xxy))$, Turing's fixed point combinator, which can express recursion. 

A danger with these beautiful but sophisticated mathematical techniques is that they might be used incorrectly, not capturing the intended behaviour of the programming constructs being defined. To eliminate this risk, one could specify the behaviour in a more natural form (so called \textit{operational semantics}) and prove the two specifications to be equivalent. This was the topic of the dissertation~\citem{gordon-evaluation} for which Mike received his PhD from the University of Edinburgh in 1973, supervised by Rod Burstall. 

Mike proved the equivalence of the denotational and operational semantics of pure LISP\@. He presented an early example of what is now called a \textit{structural} operational semantics: reduction relations defined as logical inference systems.
\begin{quote}
Mike Gordon's thesis \ldots{}\ contains a pretty rule-based operational semantics, with the environment needed to model dynamic binding incorporated in the configuration; this was the first treatment of part of a real programming language. \cite[p.\ts5]{plotkin-origins-sos}
\end{quote}
LISP presented a particular challenge due to its unusual treatment of variables. And so Mike obtained an invitation from LISP's inventor, John McCarthy, to work for a year at his Artificial Intelligence Laboratory at Stanford University.

\section{Edinburgh, Stanford and Edinburgh LCF}

The period from 1970 to 1981 set the stage for Mike's career. In 1970, when Mike began his PhD research at Edinburgh, computer science there was fragmented among rival departments. He worked in the Department of Machine Intelligence, which was part of the School of Artificial Intelligence. While he undertook research on the semantics of LISP, others in the school were working on formal logic and automated reasoning.

Formal logic is concerned with precisely specified languages along with symbols for logical connectives such as ``and'' ($\land$), ``or'' ($\lor$), ``not'' ($\neg$), ``implies'' ($\to$) and the \textit{quantifiers}: ``for all'' ($\forall$) and ``there exists'' ($\exists$). A formal calculus includes strict rules for deducing conclusions from assumptions.  \textit{First-order logic} (also known as \textit{predicate calculus}) is the simplest such system. It presupposes a fixed, non-empty universe of mathematical values (which could be numbers, sets, polygons, etc.). 

There have always been those who felt that formal logic somehow captured human reasoning. During the 1970s, many practitioners of artificial intelligence felt that if one could only automate reasoning in the predicate calculus, one could automate thought itself. (Yes, it sounds ridiculous now.) McCarthy, a leading AI pioneer, held this view strongly. Mike's first meeting with McCarthy went like this:
\begin{quote}
He went to McCarthy's office. With no preliminary, John said ``I believe everything can be done in first-order predicate calculus.''
Mike said nothing.  John got up and walked out of his office.  Soon he returned though, said ``with suitable extensions'' and he left again.%
\footnote{According to Richard Waldinger, as relayed by Bruce Anderson in an email dated 2018-04-04}
\end{quote}
So when (in 1974) Mike took up a postdoctoral position at the Stanford AI Lab, he was again working on semantics alongside people focused on formal logic. He organised a discussion group on reasoning about programs, attracting researchers from Stanford and nearby research institutes. After work, he would go home to Richard Waldinger's shared house in Palo Alto. Waldinger also worked on logic and theorem proving, at the Stanford Research Institute's Artificial Intelligence Center.

One project at the Stanford AI Lab was Stanford LCF~\cite{milner-stanford}, led by Robin Milner (FRS 1988). It has an amusing backstory. In 1969, Scott wrote a manuscript \cite{scott93} introducing a logical calculus with a rule called fixed-point induction, superseding a number of earlier techniques. (Scott's logic was quite different from Hoare's, which was concerned with program code.) Scott was concerned with pure recursive functions written in the typed $\lambda$-calculus, for which he proposed a domain-theoretic semantics. He began his paper boldly:
\begin{quote}
No matter how much wishful thinking we do, the theory of types is here to stay. There is \textit{no other way} to make sense of the foundations of mathematics.\footnote{Italics in original}	\cite[p.\ts413]{scott93}
\end{quote}
Scott was firmly committing himself to the typed  $\lambda$-calculus. But one month later, Scott made the astonishing discovery of a model for the \textit{untyped} $\lambda$-calculus. So he withheld this work from publication, and it became known to researchers only through faded Xerox copies. Working at Stanford, Milner along with Whitfield Diffie (ForMemRS 2017), Richard Weyhrauch and Malcolm Newey wrote a computer program to implement Scott's logic, which Milner named the Logic for Computable Functions or LCF\@. Milner had already left Stanford by the time Mike arrived. By 1975 they were both in Edinburgh and working together on a new version of LCF, along with Chris Wadsworth.
 
 Stanford LCF had two major limitations. Stored proofs used too much memory, and its fixed command repertoire required lengthy, repetitive sequences of steps even for elementary proofs. Milner realised that he could address both problems by providing a programmable \textit{meta\-language}, which he called ML\@. Making the prover programmable allowed users to automate any repetitive steps. Moreover, through a language concept known as \textit{abstract types}, no proofs would have to be stored. An abstract type enforces the use of a fixed set of operations; by making those operations coincide precisely with a logic's rules of inference, we could define the type of theorems. The abstraction barrier would ensure that theorems were constructed strictly according to the rules. This technique works for essentially any logic~\citem{gordon-logic-lcf}. 
 
 Edinburgh LCF was finished by 1979 \citem{mgordon79}. It introduced a simple and effective architecture for \textit{interactive}---as opposed to fully automatic---theorem proving. And far from being a mere metalanguage, ML \citem{gordon-metalanguage} was seen as a general programming language with a highly innovative design. Mike had been fully involved in these great achievements~\citem{gordon-tactics-milner}, but was already preparing to strike out on his own. He had already written what would become the standard textbook on denotational semantics~\citem{gordon-denotational}. With software verification apparently becoming a reality, Mike was the first to think seriously about verifying hardware.

 By 1979, Edinburgh's Department of Computer Science had been transformed by a crowd of new arrivals. These included Rod Burstall and Gordon Plotkin (FRS 1992), who had moved from the Department of Artificial Intelligence, as well as Robin Milner, who had arrived earlier. Hardware and systems people found themselves cheek by jowl with a great many theoreticians. Mike's friendly and modest personality allowed him to overcome resentful tribal divisions. He wanted to investigate the semantics of hardware, and that required talking to the hardware specialists. By 1981, Mike had elaborated an approach to hardware verification --- including theoretical development and fully worked out examples --- that could scale to large devices \citem{gordon-chdl,gordon-model}. He also had an invitation to join the rapidly expanding Computer Laboratory at Cambridge.
 
\section{Cambridge and the Emergence of Hardware Verification}

The first user of Edinburgh LCF was Avra Cohn. A PhD student of Milner's, she had used it to prove the correctness of an abstract compiler~\cite{cohn-phd,cohn83}. She was also Mike's wife. They had first met at Richard Waldinger's house during Mike's postdoctoral year at Stanford. Now, years later, they were sharing an office at Edinburgh. As the first LCF user, Avra influenced its design by pointing out bugs and suggesting improvements. She and Mike were already working together, a collaboration that would continue for many years. They got married in 1979, and together they brought LCF to Cambridge.

As a new University Lecturer, Mike had much to occupy him. By October 1983, he was teaching an advanced course entitled \textit{Topics in Programming Language Theory} \citem{gordon-topics}, with an ambitious syllabus: the predicate calculus, Hoare logic, the $\lambda$-calculus, automatic theorem proving using the resolution method, and logic programming. Some of the material from his course notes later found its way into his second textbook~\citem{mgordon88b}, covering programming language theory and including LISP code to implement some of the techniques.

 He also held a Science Research Council grant (jointly with Milner at Edinburgh) to continue the LCF project. Here I entered the picture, having been hired as a post-doc under this grant. I still remember Mike's kindness in meeting me at the airport and helping me take all my stuff to Cambridge.  Avra helped me to get started with LCF. She gave me her code, a bundle of utilities written in ML to help carry out LCF proofs. These included sophisticated heuristic tools based on pattern matching. It's remarkable that this code had not already been incorporated into Edinburgh LCF, which was truly a bare-bones environment. Modified and extended by myself and others, Avra's code lives on in today's systems. For I had decided to take Edinburgh LCF apart, and aided by G\'erard Huet of the Inria%
 \footnote{\textit{Institut national de recherche en informatique et en automatique}, the French national research institute for computer science.}
  lab near Paris, put it back together again. The point was to make LCF more usable and much, much faster.

Meanwhile, Mike was continuing to develop his ideas. We can trace their evolution from his 1981 Edinburgh technical report~\citem{gordon-model}. At 75 pages, this was a substantial document, not to be confused with the short conference version~\citem{gordon-chdl}. Already he was treating both \textit{combinational} devices such as adders and \textit{sequential} devices such as storage registers. Some examples were at the gate level and others were at the transistor level.

\begin{figure}[hbt]
\begin{center}
  \includegraphics[scale=0.7]{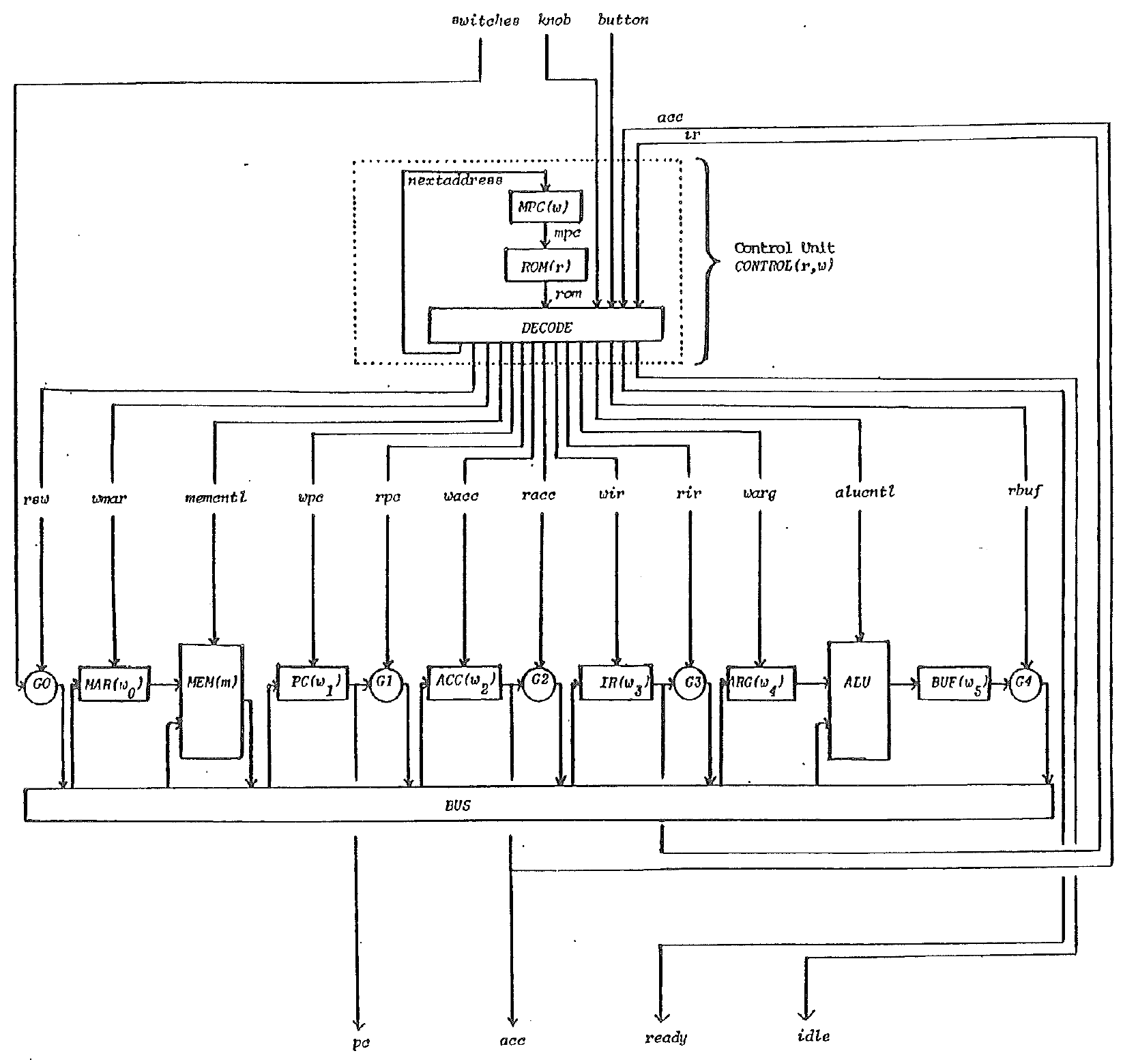}	
\end{center}
  \caption{The Gordon Computer \protect\citem{gordon-model}}\label{fig:computer}
\end{figure}

From the beginning, Mike had the ambition of scalability. He presented a simple microcoded computer (Fig.\ts\ref{fig:computer}) complete with a specification of the machine instructions and microinstructions, including a microprogram. The detailed design took up 21 pages. 

While combinational devices can easily be modelled as functions from inputs to outputs, sequential devices are trickier to formalise, as they have internal state. Mike's initial idea was to use the power of domain theory. First he defined the domain of signals $\textit{Sig}[X]$ (where $X$ is a set of wires) to denote the set of functions from $X$ to some fixed set of values. Sequential devices were also modelled as functions, incorporating the internal state as part of the result~\citem[p.\ts8]{gordon-model}:
\begin{quote}
The domain $\textit{Seq}[X;Y]$ of sequential behaviours from $X$ to~$Y$ is defined to be the least solution of the domain equation:
\[ \textit{Seq}[X;Y] = \textit{Sig}[X] \to (\textit{Sig}[Y]\times \textit{Seq}[X;Y])\]
\end{quote}
Such a behaviour maps the input~$X$ to the output~$Y$ paired with a new $\textit{Seq}[X;Y]$ (which models the possibility of an internal state change). A precursor to this technique can already be seen in his brief note on the semantics of sequential machines~\citem{gordon-machines}. For the sake of uniformity, he proposed regarding combinational devices as the degenerate case of sequential devices (with an empty internal state), so everything would involve recursive domain equations. But he was unhappy with this high-powered approach~\citem[p.\ts9]{gordon-model} and was apparently trying to use operational semantics:
\begin{quote}
The reader might wonder why we use sequential behaviours at all --- why not just work with machines?\ldots{} In fact, at various times during the development of our model, we have tried to eliminate behaviours in favour of machines, in order to avoid having to use the recursive domain equation which defines $\textit{Seq}[X;Y]$. We have never succeeded.
\end{quote}
But eventually, he did succeed, finding something even simpler than ``machines'': pure logic.

\begin{figure}
\begin{center}
  \includegraphics[scale=0.5]{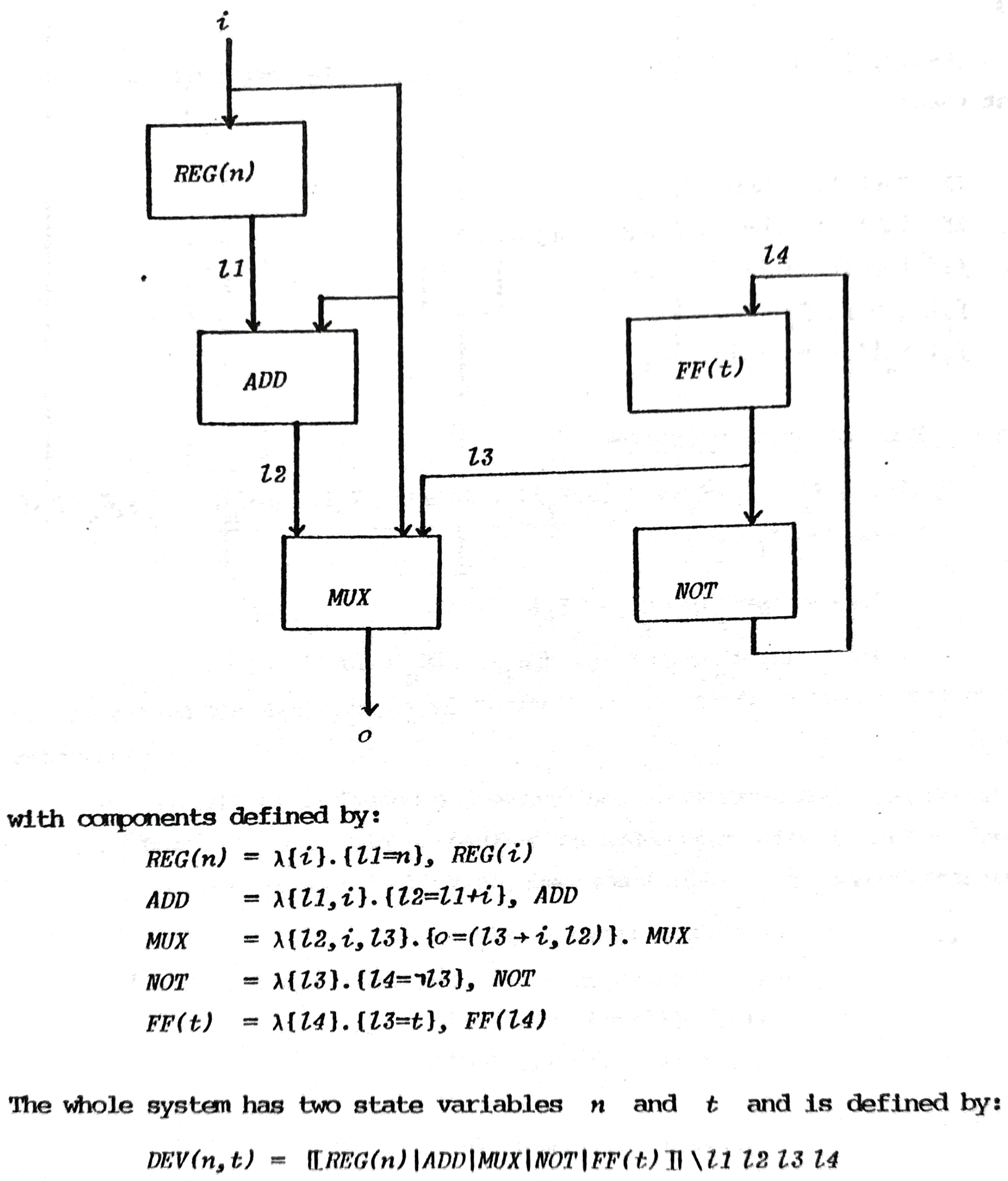}	
\end{center}
  \caption{Extract from Mike's 1981 report on hardware verification}
  \label{fig:1981}
\end{figure}

His ambition reflected the broad scope of denotational semantics and the power of domain theory. The components of a computer, including families of input lines carrying time-indexed signals, could be modelled by mathematical functions, possibly nested. It's striking to see diagrams in this early report typical of his much later work (Figs.\ts\ref{fig:computer} and~\ref{fig:1981}).  Its main ingredients were already evident, including the notational devices for connecting devices together and hiding internal wires. The mathematical underpinnings would change drastically, but the conception remained the same.

By 1983, Mike had put his ideas into practice with his Logic for Sequential Machines. He implemented this formalism on top of the Cambridge LCF code base, calling the resulting system LCF\_LSM \citem{gordon-lcflsm}. Two major changes are evident from his former work. One was the abandonment of domain theory, with its requirement that every domain had to be a partial ordering. The need to deal with the associated ``bottom'' value ($\bot$) tended to clutter proofs. Mike thought it could go away temporarily. (It never came back.) 

That led to the other major change: the replacement of functions by machines. Previously \citem{gordon-model} he had used $\textit{Seq}[X;Y]$ to denote a domain of functions, including the possibility of a state change. Now Mike had figured out how to model sequential devices without using functions, while continuing to regard a combinational device as simply a sequential device with an empty state. 

LCF\_LSM was inspired by Milner's Calculus of Communicating Systems (CCS), a mathematical model of concurrent computing \cite{milner80}. CCS is concerned with systems composed of a fixed number of processes that can send messages to each other synchronously (where the sender and receiver act at the same time) and change state. CCS includes principles for demonstrating that two apparently different systems exhibit identical behaviour. Similarly, LCF\_LSM concerns components with labelled wires that can be connected together. Wires can also be renamed or hidden. In LCF\_LSM, we can write both specifications of desired behaviour and implementations built from smaller components. We can prove that two components  have the same behaviour and prove that implementations satisfy a specification.

\begin{figure}[hbt]
\begin{center}
  \includegraphics{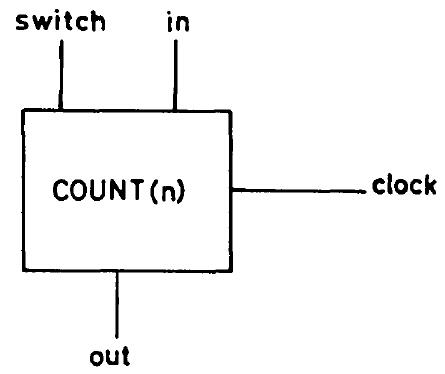}	
\end{center}
  \caption{A counter, from Gordon and Herbert \protect\citem{gordon-herbert}}\label{fig:counter}
\end{figure}

To illustrate the notation, the following formula specifies the behaviour of the counter in Fig.\ts\ref{fig:counter}. 
\begin{verbatim}
COUNT(n) == dev{switch,in,out}.{out = n}; COUNT(switch->in|n+1)	
\end{verbatim}
The device has the three wires shown (the system clock is implicit and never appears in specifications). The output line equals the counter's stored value. At each clock tick, the counter loads the value of the input line if \texttt{switch} is true and otherwise increments itself.

Great things were achieved with LCF\_LSM\@. John Herbert \citem{gordon-herbert} used it to verify a bespoke chip design for the Cambridge Fast Ring, an early local area network. Mike used it to verify his computer~\citem[p.\ts1]{gordon-computer}:
\begin{quote}
The entire specification and verification described here took several
months, but this includes some extending and debugging of LCF\_LSM
(necessary, as this was our first big example). I estimate that it would
take me two to four weeks to do another similar exercise now. The
complete proof requires several hours CPU time on a 2 megabyte
Vax750. I found it necessary to prove some of the bigger lemmas \ldots{} in batch mode overnight.
\end{quote}
This tremendous achievement demonstrated that hardware verification was becoming a reality. Nevertheless, Mike was not satisfied~\citem[p.\ts22]{gordon-lcflsm}:
\begin{quote}
The selection of rules currently included in LSM is rather ad hoc --- I have just implemented what seemed needed for the examples I have done. \ldots{} Further experimental work is needed. 
\end{quote}
Later in the report (pp.\ts37--8), he mentions the possibility of replacing LSM by some form of predicate logic.

\section{Higher-Order Logic, the HOL System and the VIPER Microprocessor}
Today Mike's wish to use ordinary logic may seem natural, but in the 1980s many people were introducing specialised formalisms. I had given myself the research goal of providing support for multiple formalisms, only to see Mike's choice of higher-order logic gradually take over the verification world. Few people favoured his choice at the time. I certainly didn't, sharing the views of most logicians:
\begin{quote}
Unlike first-order logic and some of its less baroque exten\-sions, second and higher-order logic have no coherent well-established theory; the existent material consisting merely of scattered remarks quite diverse with respect to character and origin. \cite[p.\ts241]{van-benthem-higher}
\end{quote}
First-order logic was also strongly preferred by many researchers in artificial intelligence, such as McCarthy at Stanford, as we have seen. And yet, higher-order logic could be seen as a return to tradition:
\begin{quote}
The logics considered from 1879 to 1923 \ldots\ were generally richer than first-order logic [and] \ldots\ at least as rich as second-order logic \ldots{} It was in Skolem's work on set theory (1923) that first-order logic was first proposed as all of logic and that set theory was first formulated within first-order logic. \cite[p.\ts127]{moore-emergence}
\end{quote}
The difference between these ``orders'' of logic concerns their treatments of sets and functions. Recall that the symbol $\forall$ (the universal quantifier) means ``for all'' and we can write statements like $\forall x y.\, x+y = y+x$ to assert the commutativity of addition. Here, $x$ and $y$ presumably range over numbers of some sort. But consider the following logical formula:
\begin{equation}
\forall P.\, [P(\textit{True}) \land P(\textit{False}) \to \forall x.\, P(x)].
\label{eqn:TF}
\end{equation}
The universally quantified variable, $P$, is a predicate, and $P(x)$ is a formula.  But quantification over predicates is forbidden in first-order logic. First-order logic allows quantification only over some fixed domain of individuals; second-order logic also allows quantification over functions and predicates defined on individuals; higher-order logic allows quantification over arbitrary functions and predicates whose arguments may themselves be other functions and predicates.

Higher-order logic includes a type system to govern all this. For first-order logic there is no need, as all variables range over individuals and it is not essential to introduce different sorts of individuals, although this is sometimes done anyway. With higher-order logic, Church~\cite{church40} used the following types:%
\footnote{Church used a different syntax, nearly incomprehensible to modern eyes.}
\begin{itemize}
	\item $\iota$, the type of individuals
	\item $o$, the type of the truth values \textit{True} and \textit{False}
	\item $\sigma\to\tau$, the type of functions from $\sigma$ to~$\tau$
\end{itemize}
These include as a special case $\sigma\to o$, the type of predicates on type~$\sigma$. For formula~(\ref{eqn:TF}) to make sense, the variable $P$ must have type $o\to o$ and $x$ must have type~$o$. Higher-order logic is an extension of Church's typed $\lambda$-calculus.
 
 Mike introduced higher-order logic to the verification world in 1986~\citem{mgordon86}, sketching its syntax and semantics. He presented examples including an inverter, a full adder (implemented in terms of transistors) and a sequential multiplier. The state in a sequential device is modelled by taking the values on wires to be functions of time, indexed by integers. Then the output of a device at time $t+1$ can be related to its input at time~$t$. Mike credited Ben Moszkowski with ideas for reasoning about timing properties. Credit for the suggestion of higher-order logic went to Keith Hanna \cite{hanna-specification}, who later decided to try his luck with more advanced type theories. But Mike's simple choice was the right one.
 
\begin{figure}[hbt]
\begin{center}
  \includegraphics[scale=0.6]{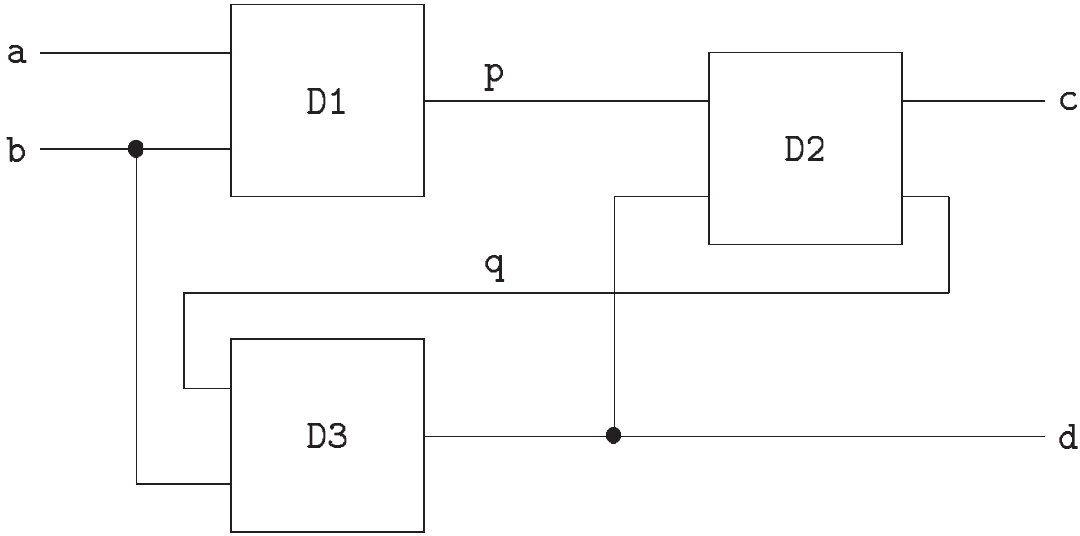}	
\end{center}
  \caption{Representing circuit structure with predicates \protect\citem[p.\ts157]{mgordon86}}\label{fig:predicates}
\end{figure}

 Mike's paper contains the definitive enunciation of the approach of representing hardware devices by relations or predicates. Recall that  device behaviours were given first by recursive domain equations \citem{gordon-model} and then by dedicated terms \citem{gordon-lcflsm}. But with higher-order logic, the behaviour of a device~$D$ is simply a relation over $D$'s external lines, with no distinction between inputs and outputs. Devices are connected together by equating the corresponding lines. Wires are hidden from the outside by existential quantification: mathematically, this is the composition of relations. For example, the formula%
 \footnote{There is an error in this formula in Mike's original paper~\citem{mgordon86}.}
 \[ \exists p\, q.\,D_1(a,b,p)\land D_2(p,d,c,q)\land D_3(q,b,d) \]
 represents the device shown in Fig.\ts\ref{fig:predicates}. Two standard logical symbols, $\exists$ and~$\land$, have replaced the special notation we saw in the last line of Fig.\ts\ref{fig:1981}.

\begin{figure}[hbt]
\begin{center}
\includegraphics[scale=0.8]{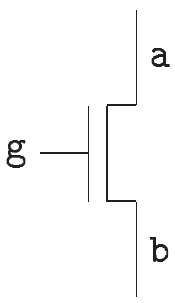}		
\end{center}
  \caption{An n-type transistor \protect\citem[p.\ts159]{mgordon86}}\label{fig:ntrans}
\end{figure}

The relational approach is the right way to model individual transistors. Terminals $a$ and $b$ are neither inputs nor outputs, but are the terminals of a switch, controlled by~$g$, the gate (Fig.\ts\ref{fig:ntrans}). 
Mike treated an inverter containing two transistors. Note that the power and ground are viewed as explicit components, connected to the transistors by internal wires, $p1$ and~$p2$. Later in the paper, Mike treats a full adder consisting of 24 transistors. He credits this example to Inder Dhingra and comments, ``Such a proof would be difficult with the usual representation of combinational circuits as boolean functions. Relations rather than functions are needed to model bidirectionality.'' \citem[p.\ts162]{mgordon86}
\begin{figure}[hbt]
\begin{center}
  \includegraphics[scale=0.7]{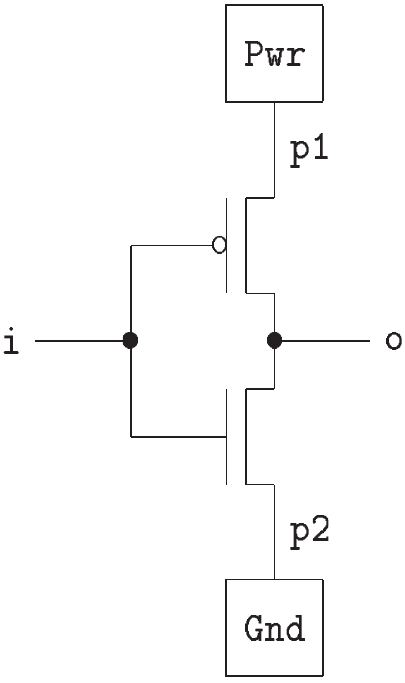}	
\end{center}
  \caption{A CMOS inverter \protect\citem[p.\ts158]{mgordon86}}\label{fig:inverter}
\end{figure}

The methodology for verifying such a device is simplicity itself and scales all the way from this inverter to a full-sized computer. You define two predicates, say \texttt{INVERTER} (describing the desired behaviour of the inverter) and \texttt{INVERTER\_IMP} (describing an implementation in terms of smaller components, as in Fig.\ts\ref{fig:inverter}). Those smaller components will typically be regarded abstractly; there is no need to go all the way down to the transistor level. Then you prove that \texttt{INVERTER\_IMP}$(i,o)$ implies \texttt{INVERTER}$(i,o)$ for all $i$ and~$o$. This states that every configuration of values on the wires permitted by the implementation is also permitted by the specification.

Some weaknesses of the methodology are also clear. One is that electronic issues such as gate delays and voltage levels are abstracted away. This approach will not tell you that one output is trying to drive too many inputs or that a combinational circuit is too slow. It is a general limitation of mathematical models that they can never capture the real world in full. 

A specific limitation of this approach is that there exists one implementation that satisfies all specifications. Simply connect power to ground; that is formalised as $1=0$, which can prove anything. Nobody would do this on purpose, but a design could accidentally short circuit for certain combinations of inputs. The specification would be satisfied but the implementation would burn. One solution to this difficulty is to prove the converse of the implication above (every behaviour allowed by the specification is satisfied by the implementation), but this is not always possible: most specifications allow some diversity of behaviours. Other measures can be used to check the sanity of the implementation.

Once again, Mike had the task of building a theorem prover, starting with the Cambridge LCF base and creating the world's first interactive implementation of higher-order logic. Avra Cohn was again the first user and, along with Mike, verified a counter circuit~\citem{cohn-counter,cohn-counter-tr}.%
\footnote{The technical report \citem{cohn-counter-tr}
 contains the full HOL proof, some 30 pages of code.}
 This was a pilot study towards the first landmark HOL proof: the VIPER 32-bit microprocessor~\cite{cullyer-viper}. The counter, which originated with the UK's Royal Signals and Radar Establishment (RSRE), comprises nine flip-flops and a couple of dozen gates including the counter logic. A  complication of the design is that one can request either a single or a double count; the latter is implemented by calling the increment logic twice, so the machine has a two-bit control state and its timing is not uniform. The verification requires reasoning about temporal properties of the circuit.
 
 The verification of the VIPER microprocessor was the first proof of its kind, establishing HOL as a verification platform for realistic hardware. Yet again, this was the work of Avra Cohn. VIPER was designed by RSRE for military purposes, hence the interest in verification; it was specified in a series of levels, from abstract to concrete. Cohn verified the equivalence of the first two levels~\cite{cohn88}, and later, the second pair of levels~\cite{cohn89a}.
 
 Overshadowing these achievements was a controversy over what Cohn had actually accomplished~\cite{mackenzie-viper}. Exasperated by exaggerations of her work in marketing material, she wrote a paper \cite{cohn89b} pointing out the inherent limitations of her work in particular and hardware verification in general. She had indeed verified a major part of the VIPER design but not down to the gate level, and the specification omitted some important operating modes. More fundamentally, ``verification involves a pair of models that bear an uncheckable and possibly imperfect relation to the intended design and to the actual device'' \cite[p.\ts131--2]{cohn89b}. In other words, both the designer's objectives and the device's physical manifestation lie beyond the scope of formal verification.

\section{The Golden Age of HOL}

The name of Mike's new prover, HOL88, marks 1988 as the official start of the higher-order logic era~\citem{mgordon88a}. The achievements reported above had already been attracting a steady stream of PhD students. Graham Birtwistle and Jeff Joyce~\citem{joyce-proving} used HOL88 to verify a simplified version of the Gordon Computer, which they called Tamarack.%
\footnote{Recently, Thomas T\"urk got a version of this old proof working on the latest version of HOL. It now runs in a couple of seconds.}
Tom Melham developed a comprehensive package for defining recursive data structures~\cite{melham89}, such as lists and trees; with Mike, he wrote the first HOL manual~\citem{mgordon-hol}. And there was much more. International meetings on hardware verification were dominated by work done using HOL88 \cite{birtwistle88,birtwistle89}. In 1991, Sara Kalvala compiled a snapshot of HOL activity around the world, listing over eighty diverse projects~\cite{kalvala-hol}. 

By this time, HOL88 was being supplanted by Konrad Slind's faster HOL90, which eventually became today's HOL4~\cite{slind-hol4}. Other systems inspired by HOL88 include John Harrison's HOL Light~\cite{hol-light-tutorial}. In the USA, researchers chose an extended form of higher-order logic as the basis for their Prototype Verification System (PVS) \cite{PVS-CADE92}. With my own verification tool (Isabelle), I would continue to push first-order logic and set theory as a basis for verification until the late 90s, when the dominance of higher-order logic became overwhelming. The other major formalism for verification is dependent type theory, exemplified by Coq~\cite{coq91}, which is a powerful extension of higher-order logic. 

Mike was elected to the Royal Society in 1994, the year when the risk posed by hardware defects burst into public view. A floating-point division error in the Pentium processor forced Intel to recall millions of chips at a cost of \$475 million~\cite{nicely-pentium-fdiv}. Until that date, many theorem provers did not even support negative numbers; it was suddenly urgent to deal with floating-point arithmetic and numerical algorithms. Harrison tackled this \cite{harrison94}; he went on to accomplish great things in formalised mathematics, including verifying a floating-point exponential function \cite{harrison-exp} and (much later) playing a major role in verifying the celebrated Kepler conjecture~\cite{hales-formal-Kepler}.

Another landmark was the verification of probabilistic algorithms, which exploit randomness. They can achieve great efficiency, but their result is only guaranteed to be correct with a certain probability, e.g.\ of the form $1-2^{-n}$. To verify such an algorithm means to show that the probability of an error is no worse than the specification. Joe Hurd formalised enough measure theory to verify a variety of probabilistic algorithms~\cite{hurd-primality}. Harrison and Hurd's work led to the substantial libraries of analysis found in many of today's verification systems. They are just two of Mike's many students who did great things in HOL's golden age.

\section{Software Verification, ARM6 and Verified Compilers}

Mike's most far-reaching project was his collaboration with Graham Birt\-wistle to verify a modern processor. By the year 2000, several processors had been formally verified, but none were full-scale commercial designs containing advanced features such as instruction pipelining. The project involved working with ARM, whose processors are found in billions of mobile phones around the world. Anthony Fox, working at Cambridge, verified the ARM6 processor. This work yielded a complete specification of the ARM6's instruction set architecture. Other researchers built projects upon that, aimed at verifying machine-language code~\citem{fox-specification}. But to tell this story properly, we need to go back to the 1980s.

With HOL, Mike introduced a strict treatment of definitions: a new constant~$c$ could be introduced only by asserting $c=t$, where $t$ is a $\lambda$-term not mentioning~$c$ and without free variables. While axioms can lead to contradictions, definitions are conservative. Mike also introduced a principle for declaring new types as non-empty subsets of other types~\citem{mgordon-history}.  \textit{Recursive} definitions would require explicit fixed point constructions, though these would soon be automated using ML \cite{melham89}. The HOL group may have had Bertrand Russell in mind \cite[p.\ts71]{russell-mathematical}:  
 \begin{quote}
The method of `postulating' what we want has many advantages; they are the same as the advantages of theft over honest toil.
 \end{quote}
 Russell was referring to the tedious construction of the real numbers from the rationals using Dedekind cuts, which was formalised by Harrison~\cite{harrison94}. While other verification groups preferred theft, Mike and his students were firmly committed to rigour. 

In the 1970s, Mike had chosen hardware verification because software verification seemed likely to be solved soon. But that clearly wasn't happening (it still hasn't), and already in 1988, Mike was thinking about using HOL to verify software. 
\begin{quote}
The work described here is part of a long term project on verifying combined hardware/software systems by mechanized formal proof. \citem[p.\ts3]{gordon-mechanizing}
\end{quote}
This eventually led to intensive research into techniques of verifying software, in ML-like languages and machine language, right down to the bit level. 

The dominant approach to software verification, Hoare logic~\cite{hoare-axiomatic}, concerned triples of the form
\[ \{P\}\;S\;\{Q\} \]
where $S$ was an executable statement, $P$ was the precondition and $Q$ was the postcondition. This Hoare triple asserted that $Q$ would hold after the execution of $S$ provided $P$ held beforehand and the execution terminated. Hoare logic allowed clear, natural proofs, but many difficulties soon manifested themselves. It assumed that the Boolean expressions of the programming language could be identified with the quantifier-free formulas of the assertion language in which $P$ and $Q$ were written. But Boolean expressions are executable and subject to all the ambiguities and complexities that make semantics necessary in the first place. Many verification systems based on Hoare logic were of doubtful correctness or required users to assume many axioms. 

Mike decided to implement Hoare logic upon the sound and expressive platform of HOL\@. His innovation \citem{gordon-mechanizing} was to define a simple programming language by a formal operational semantics; the Hoare-style rules would then be derived, not simply asserted. Following his definitional approach, there would be no axioms. Through the power of ML --- a modified pretty-printer disguising all the machinery --- users would be given the illusion that they were working in Hoare logic.

This was the first example of what is now called a \textit{shallow embedding}: a formalism (here Hoare logic) is not defined in HOL but merely simulated, yielding a convenient proof environment for that formalism. If instead we define the formalism inductively as a mathematical object, then we have a \textit{deep embedding}. The formalism's metatheory can easily be developed, but conducting derivations within the formalism will be painful. Over the years, many assertion languages would be implemented in HOL and other systems using one or the other approach~\citem{bowen-gordon-shallow}. Hoare-style precondition/postcondition calculi remained a favourite. These techniques were well understood by the year 2000, when the ARM6 verification project commenced.

This landmark project, jointly between the universities of Cambridge and Leeds, was funded by the EPSRC\@. Birtwistle at Leeds would specify the instruction set architecture (ISA) and the processor implementation;%
\footnote{The ISA describes the computer as a machine language programmer sees it. The  implementation is in terms of memory, registers and an arithmetic/logic unit (ALU).}
 Mike at Cambridge would formalise and verify these specifications using HOL4. Anthony Fox, a postdoc of Mike's, undertook the Cambridge task and took about a year to prove that a model of the ARM6 processor correctly implemented the corresponding ISA\@. Fox went on to specify  other ARM instruction sets, and independently, other researchers formalised the x86 and PowerPC\@. These exceptionally detailed ISA specifications (and associated tools) formed a resource that would be widely used.
 
  With Magnus Myreen, a new PhD student, Mike decided to verify machine code programs. Prior work on verifying machine code was frustrated by the \textit{frame problem}: the need to state explicitly which parts of the machine state were left unchanged. (When you flush the toilet, you don't wonder whether your car doors will unlock.)
  
  A formalism known as separation logic \cite{reynolds-separation-logic} had been proposed to deal with the frame problem, and Mike suggested adapting those ideas to higher-order logic. Myreen developed techniques to generate Hoare-style assertions for each machine instruction while specifying only which parts of the state changed~\citem{myreen-fox-hoare,myreen-hoare}. He was then able to make a \textit{decompiler}: to translate a string of machine instructions into a mathematical function expressing the state transformation, the equivalence automatically verified in HOL4~\citem{myreen-machine-code,myreen-function-extraction}. To crown it all, verified decompilation provided a means of verifying the result of \textit{compilation}: the translation of source code to machine code. Myreen's technology allowed him to create verified LISP interpreters in three different machine languages~\citem{myreen-LISP}. Myreen's PhD thesis won the British Computer Society's Distinguished Dissertation Award in 2010. His choice of LISP echoes Mike's own PhD thesis~\citem{gordon-evaluation}.

These outstanding results attracted substantial follow-up funding. One of the most striking outcomes is CakeML, a version of the ML language implemented as a mathematical function in HOL \cite{kumar-cakeml-verified}. Ramama Kumar et al.\ followed a ``bootstrapping'' procedure, initially using HOL itself to translate fragments of CakeML into binary code; they thus obtained a usable compiler that has been proven to generate correct binary code. This solves the chicken and egg problem of compiler correctness: if you verify a compiler that is written in a high-level language, what compiler do you use to translate it correctly into binary? Mike's students and colleagues could not resist the temptation to apply these techniques to HOL itself~\cite{kumar-self-formalisation}. And so another of Mike's students was honoured: Kumar won the ACM SIGPLAN Doctoral Dissertation Award for 2017.

\section{Legacy}

The verification world of today is substantially shaped by Mike's work. Conferences for HOL users have been held annually since 1988, now broadened to related systems under the name Interactive Theorem Proving (ITP). The leading interactive theorem provers follow the LCF approach, are implemented in some version of ML, and support higher-order logic or something stronger. Hardware verification is widely used in industry, while academic research continues apace. 

Mike was always keenly interested in all these developments. He worked on many projects connected with hardware description languages, interoperability of verification tools and other technologies. He was fully aware of rival methods, including model checking (to verify system properties by enumeration of finite but large state spaces) and binary decision diagrams (BDDs: graph-based data structures capable of manipulating extremely large propositional formulas efficiently). He found an ingenious way of combining BDDs with HOL~\citem{gordon-bdd,gordon-puzzletool}. He admired the hardware verification research of the University of Texas at Austin using ACL2 --- a theorem prover based on an utterly different design from HOL's --- and worked to link up that prover with HOL~\citem{gordon-embedding-acl2}, combining their complementary strengths.

Although Mike rejected engineering as a degree course, it's clear that he wanted to make an impact on the world. By talking to real hardware designers, he learnt about their practices and problems. He devoted his career to finding realistic solutions. Ironically, although his decision to tackle hardware may have been prompted by a feeling that software was being solved, software developers have generally been uninterested in verification: software can always be patched, and the industry is protected by sweeping warranty disclaimers. However, hardware is not fully solved: the complexity of modern processor designs still makes complete verification unaffordable. Only a few critical components get formal scrutiny.

Much more could be written. Many of Mike's other students accomplished great things and found prominent positions in academia or industry. Mike had a keen interest in computational linguistics: he obtained a Masters degree in linguistics from Cambridge in 1974, and engaged in sponsored research along with Stephen Pulman on applications of higher-order logic to the semantics of natural language. Mike had many teaching and administrative responsibilities, including his role in the planning of the William Gates Building, which now houses the Department and opened in 2001, and his many duties as Deputy Head of Department. 

Then there is his personal life. Avra, his wife, eventually retired from active research to bring up their two sons, Katriel and Reuben. She and Mike continued to discuss verification at home. Both of their sons went on to do PhDs in computing: Katriel in cybersecurity at Oxford, Reuben in computational linguistics at Stanford. Somehow this completes the circle.

 Mike will be remembered for his kindness and modesty --- always eager to confess his failings while concealing his triumphs --- and his gentle sense of humour.

Additional information on the history of this period has been written by Mike himself~\citem{mgordon-history,gordon-tactics-milner} and by his colleagues~\cite{harrison-history,paulson-computational}.

\paragraph{Acknowledgements}

Avra Cohn and Katriel Cohn-Gordon answered many questions and made unique manuscripts available. Mike's former colleagues, students and others supplied valuable tidbits of information and insightful comments. These include Bruce Anderson, Jasmin Blanchette, Jon Crowcroft, Warren Hunt, Sara Kalvala, Simon Laughlin, Joe Leslie-Hurd, Stephen Levinson, Magnus Myreen, Michael Norrish, Gordon Plotkin, Lee Smith, Terence Moore and Richard Waldinger. 

\paragraph*{Data accessibility.}
No experimental data involved. 

\paragraph*{Competing interests.}
Not applicable.

\paragraph*{Funding.}
Much of the research reported here was supported by the EPSRC or its predecessors, or EU funding agencies, going back 40 years. 

\paragraph*{Short bio of author.}
Lawrence Paulson FRS is Professor of Computational Logic at the University of Cambridge, where he has held established positions since 1983. He has written over 100 refereed conference and journal papers as well as four books. In the 1980s, he worked with Mike Gordon on further development of the LCF proof assistant, which became the foundation of Gordon's LCF\_LSM and HOL systems. He introduced the popular Isabelle theorem proving environment in 1986, and made contributions to the verification of cryptographic protocols, the formalisation of mathematics, automated theorem proving technology, and other fields. He achieved a formal analysis of the ubiquitous TLS protocol, which is used to secure online shopping, and the formal verification of {G\"odel}'s second incompleteness theorem. In 2008, he introduced MetiTarski, an automatic theorem prover for real-valued functions such as logarithms and exponentials. He has the honorary title of Distinguished Affiliated Professor from the Technical University of Munich and is a Fellow of ACM as well as the Royal Society. He holds a PhD in Computer Science from Stanford University, and a BS in Mathematics from the California Institute of Technology.

\paragraph*{NOTE TO EDITOR.} The Computer Laboratory has undertaken to ensure the continued validity of the URLs for Mike's autobiographical webpages (dated 2017).

\renewcommand{\refname}{References to Other Authors}
\bibliographystyle{abbrv}
\bibliography{string,gordon,atp,general,isabelle,theory,funprog,crossref}
\bibliographystylem{plainyear}
\bibliographym{string,gordon,atp,general,isabelle,theory,funprog,crossref}
\end{document}